\documentclass[11pt]{article}
\usepackage{floatflt}
\usepackage{amsmath}
\usepackage{graphicx}
\usepackage{amssymb}

\makeatletter

\providecommand{\tabularnewline}{\\}

\usepackage{moriond}
\usepackage{graphicx}
\usepackage{color}
\usepackage{feynmp}  
\usepackage{pst-ghsb}
\usepackage{psfrag}

\usepackage{fancybox}

\newcommand{\mygreen}{\color[rgb]{0,0,0}}
\newcommand{\myred}{\color[rgb]{0,0,0}}
\newcommand{\myblue}{\color[rgb]{0,0,0}}

\newcommand{\diag}{\textsf{diag}}

\newcommand{\im}{\mathrm{Im}}

\def\slashchar#1{\setbox0=\hbox{#1}
   \dimen0=\wd0
   \setbox1=\hbox{$\diagup$}
   \dimen1=\wd1
   \ifdim\dimen0>\dimen1\hbox{
      \rlap{\hbox to \dimen0{\hfil\box1\hfil}}
      \box0}
   \else\hbox{
      \rlap{\hbox to \dimen1{\hfil\box0\hfil}}
      \box1}
   \fi}

\def\lessim{\hbox{\lower.8ex\rlap{$\sim$}\lower-.2ex\hbox{$<$}}}

\makeatother
\begin{document}
\begin{fmffile}{su2anomalyproc}

\vspace*{4cm}

\title{LEPTOGENESIS: A LINK BETWEEN \\
 THE MATTER-ANTIMATTER ASYMMETRY AND \\
 NEUTRINO PHYSICS}

\author{J. ORLOFF}
\address{Laboratoire de Physique Corpusculaire, Universit\'{e} Blaise Pascal,
24 rue des Landais, F-63177 Aubi\`{e}re}

\maketitle
\abstracts{We review the experimental evidence for a net baryon density in cosmology, and the theoretical mechanism for producing it, called leptogenesis, which relies on the creation of a lepton asymmetry at an intermediate step.  The naturality of this mechanism and its possible relations with neutrino oscillations are outlined. }

\section{Facts and Fancy about the Matter Asymmetry}

Matter is so tightly connected to our everyday experience, that the
fascinating prediction of anti-matter in Dirac's theory first raised
skepticism, which turned into solid confidence after discovery of
the positron. Clearly, anti-matter on earth exists only briefly after
the high-energy collisions we use to study fundamental interactions.
At first sight, this fugacity \emph{seems} a natural explanation for
the absence of anti-matter around us. In fact, it is rather a consequence
of the domination of matter which we tend to take for granted by a
kind of anthropic argument. Quantitatively explaining this domination
(or asymmetry) of protons is the purpose of \emph{baryogenesis}\cite{Turner:1981ha}.
Let us first see how to quantify this baryon asymmetry.

Even when turning off accelerators on earth, we can collect some $10^{-4}$
anti-proton for every proton in the cosmic rays that penetrate the
upper atmosphere after a $10^{8}$ years erratic journey since their
production by supernova explosions in our galactic disc. This however
does not constitute an evidence for a $10^{-4}$ anti-supernova fraction.
Indeed, the interstellar dust is dense enough to play the role of
fixed target intercepting a small fraction of the primary cosmic proton
flux and producing the observed secondary anti-protons. This fraction
can be cross-checked\cite{Maurin:2002ua} against the amount of gamma
rays produced by the same collisions. The observed anti-proton fraction
$\bar{\mathrm{p}}/\mathrm{p}\approx10^{-4}$ is thus only an upper
bound on the natural anti-matter fraction. Incidentally, a tighter
bound of this type can be obtained by considering heavier nuclei like
deuterium, for which\cite{Boezio:1998vc,Ambriola:2002de} $\mathrm{D/p}\approx1/60$,
while we expect\cite{Chardonnet:1997dv,Duperray:2002pj} $\mathrm{\bar{D}/p}\approx10^{-8}$.
We can thus assume that in our galaxy, like on earth, there is a total
domination of matter over anti-matter, which disappeared by annihilating
with neighboring matter.

\begin{figure}
\noindent \includegraphics[%
  width=0.55\textwidth]{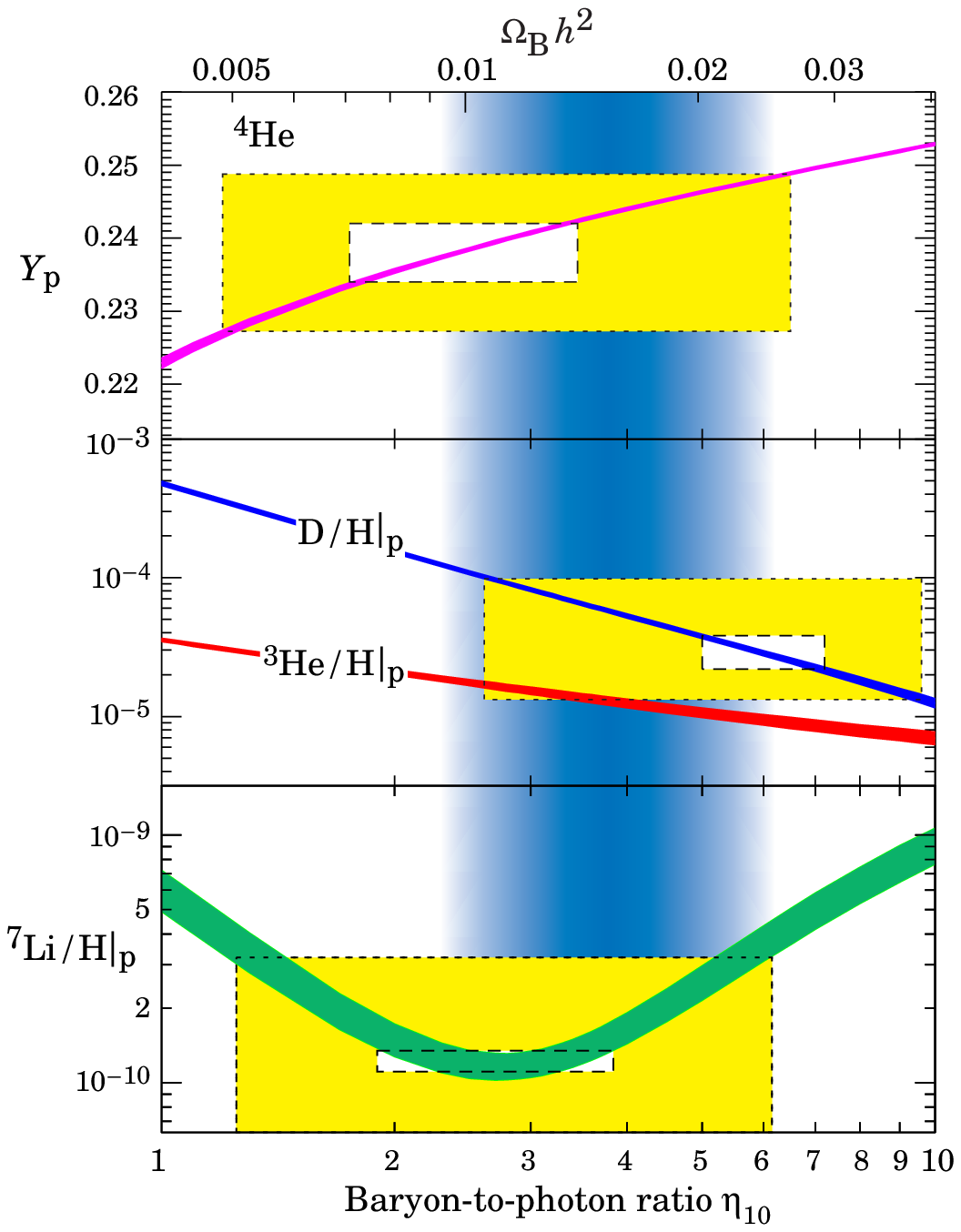}\hfill{}\begin{minipage}[b]{0.40\textwidth}%
\noindent \begin{center}\includegraphics[%
  bb=310bp 443bp 586bp 699bp,
  clip,
  width=1.0\textwidth,
  keepaspectratio]{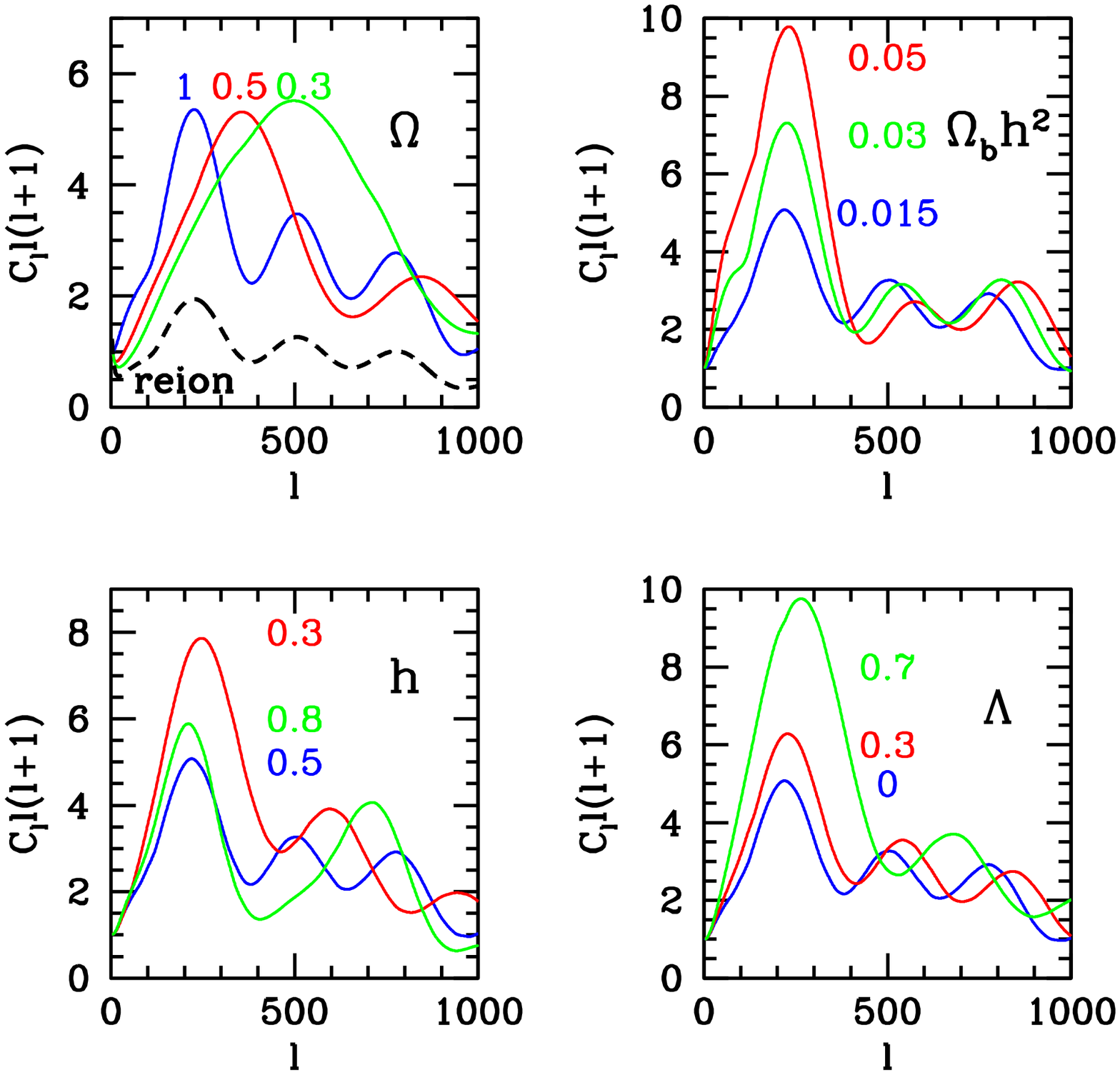}\\
\psfrag{0}[ct][ct]{0} \psfrag{5}[ct][ct]{5} \psfrag{10}[ct][ct]{10} \psfrag{15}[ct][ct]{15} \psfrag{0}[r][r]{0} \psfrag{0.2}[r][r]{0.2} \psfrag{0.4}[r][r]{0.4} \psfrag{0.6}[r][r]{0.6} \psfrag{0.8}[r][r]{0.8} \psfrag{1}[r][r]{1} \psfrag{h10}[l][l]{$\eta_{10}$} \psfrag{Likelihood}[cb][cb]{{\raisebox{2ex}{Likelihood}}} \psfrag{BBN}[l][l]{\color[rgb]{0,0,1} BBN} \psfrag{CMB}[l][l]{\color[rgb]{1,0,0} CMB} \psfrag{SNIa}[r][r]{\color[rgb]{1,0,1} SN1a} \psfrag{Latest WMAP result}[l][l]{\color[rgb]{1,0,0}WMAP 1$\sigma$} \includegraphics[%
  bb=26bp 20bp 526bp 402bp,
  clip,
  width=1.0\textwidth,
  keepaspectratio]{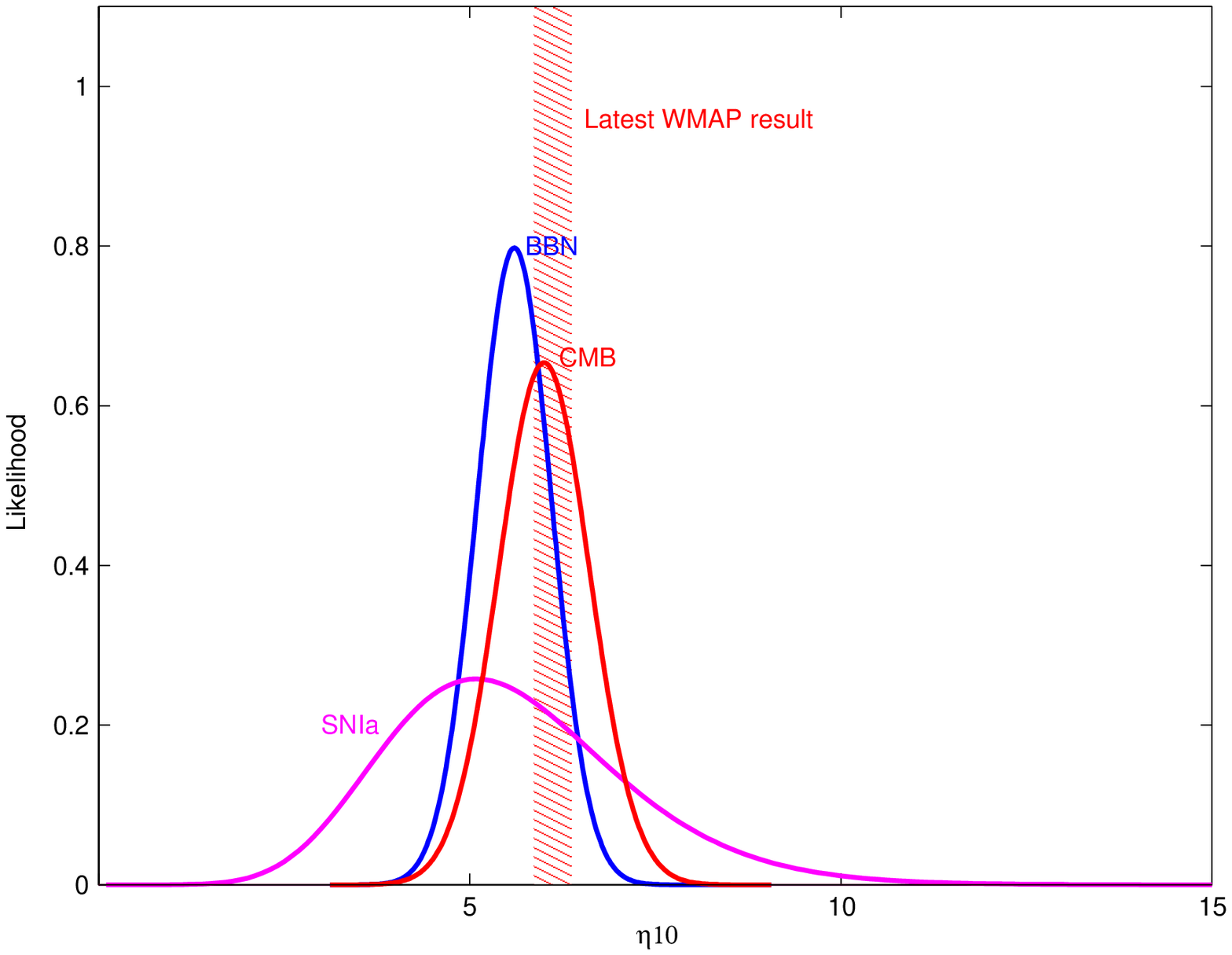}\end{center}\end{minipage}%

\caption{\label{cap:eta10}Sensitivity of cosmological observables to $\eta_{10}\doteq10^{10}\eta_{B}=274\Omega_{B}h^{2}$:
relative abundances of light nuclei produced by Big Bang Nucleosynthesis
(left); acoustic peaks in the Cosmic Microwave Background temperature
inhomogeneities (upper right); all combined with the baryon fraction
in X-emitting clusters and dark matter cosmological density induced
from Supernova Ia luminosities (lower right). }
\end{figure}

To quantify the chances of having anything surviving these annihilations,
we need a dimensionless number summarizing the baryon asymmetry of
the universe. One could take the ratio of the net baryon number density
$n_{B}\doteq(n_{\mathrm{p}}-n_{\bar{\mathrm{p}}})$ divided by $(n_{\mathrm{p}}+n_{\bar{\mathrm{p}}})$
as is often done in particle physics. However, it is more useful to
divide by the entropy density, and define $Y_{B}\doteq n_{B}/s$ which
is an invariant if 1) baryon number is conserved and 2) the expansion
of the universe is slow and adiabatic enough to avoid irreversible
entropy creation in a comoving volume. The baryon to photon ratio
$\eta_{B}\doteq n_{B}/n_{\gamma}\approx7Y_{B}$ is only equivalent
at low temperatures%
\footnote{Explicit values of $\eta_{B}$ found in the literature always refer
to this present day limit, even when extracted from physics at much
earlier times where the relation to the constant $Y_{B}$ is different.%
} $T<m_{e}$ when photons and neutrinos dominate the entropy density,
electrons and all other particles being non relativistic. 

It is difficult to evaluate $\eta_{B}$ by direct observation, because
baryons tend to clump together gravitationally, while photons don't.
To get indirect handles on $\eta_{B}$, we thus need to identify physical
processes sensitive to the baryon density $n_{B}$ averaged on cosmological
distances, or occurring before the formation of structures like galaxies.
A classical example is Big Bang Nucleosynthesis, \emph{i.e.} the making
of light nuclei in the primeval plasma before star formation. Increasing
the baryon density reduces the entropy price to pay for keeping nucleons
together in a nucleus (instead of letting them fill space) and thus
increases the $\mathrm{^{4}He}/\mathrm{p}$ ratio\cite{PDG02,Burles:2000zk}
as apparent in the left figure~\ref{cap:eta10}. However, the primordial
$^{4}\mathrm{He}$ abundance is not very sensitive, and is further
easily contaminated by Helium later formed in stars. More interesting
is the Deuterium, whose much lower binding energy per nucleon allows
to understand both its very low production in stars, and its decreasing
primordial abundance with increasing $\eta_{B}$. The ratio of these
primordial abundances to hydrogen are extracted from interstellar
clouds absorption lines into the emission of $z=0.1\to3.5$ distant
quasars. There is some tension between the Helium and Deuterium preferred
values, the latter being more trusted and giving baryon to photon
ratios around \[
\eta_{B}\doteq\frac{n_{B}}{n_{\gamma}}\approx5.6\times10^{-10}\doteq\eta_{10}\times10^{-10}.\]

Another cosmological observable that feels the baryon density is the
Cosmic Microwave Background radiation, whose temperature fluctuations
reflect the way preexisting density perturbations start oscillating
when they enter the horizon at the recombination temperature $T_{rec}\approx0.1\mathrm{eV}$
where atoms get formed. The RMS spherical harmonics coefficients $C_{l}$
of the temperature distribution display so called {}``acoustic peaks''
at peculiar values of the inverse angular scale $l$ (see the WMAP
report at this conference). Since $m_{\mathrm{p}^{+}}\gg m_{\mathrm{e}^{-}}$,
the effect of protons self-gravity is not neutral, and tends to enhance
the compression peaks ($1^{st}$, $3^{rd}$,\ldots{}) and suppress
the expansion peaks ($2^{d}$,\ldots{}). Increasing the baryon (and
electron) density also decreases the sound velocity in the plasma,
which separates the peaks from each other. These effects are illustrated
in the upper right figure~\ref{cap:eta10}, where all parameters
but the baryonic density $\eta_{10}=274\Omega_{B}h^{2}$ are fixed
at some typical values\cite{Kamionkowski:1999qc}.

A last handle on $\eta_{B}$ is the baryon fraction deduced from the
X ray emission of large clusters. Analyses of the temperature profiles
cannot be understood in terms of baryons alone, and have long been
an argument for dark matter on scales larger than the galactic flat
rotation curves. On such large scales, it can be argued that the gravitational
accretion of baryons and dark matter are similar. Knowing then the
cosmological baryon to dark matter fraction, the baryon density is
accessible through an estimate of the dark matter density, for instance
by combining the CMB and Supernova Ia data.

These three determinations are combined\cite{Steigman:2002qv} in
the lower-right figure~\ref{cap:eta10}. The concordance which emerges
is a fascinating confirmation of the adiabatic invariance of $Y_{B}$:
all the way from the nucleosynthesis temperature where we find $\eta_{10}(T_{BBN}\approx1\mathrm{MeV})=5.6\pm0.5$
from Deuterium abundance alone, to essentially today where the X clusters
baryon fraction give $\eta_{10}(T_{SNIa}\approx1\mathrm{meV})=5.1\pm1.6$,
passing through the recombination where the CMB fluctuations give
$\eta_{10}(T_{rec}\approx0.1\mathrm{eV})=6.0\pm0.6$, we find reasonable
agreement over 9 orders of magnitude in temperature or comoving volume
size. We also show the latest WMAP results reported at this conference
$\eta_{10}=6.1_{-.2}^{+.3}$ as a hatched band. We should however
mention that the quoted errors are given for all other parameters
fixed at their global $\chi^{2}$-minimizing values; minimizing $\chi^{2}$
for each $\eta_{10}$ would result in wider error bars because of
correlations, in particular with $\Omega_{DM}$.

After reviewing some quantitative facts about the baryon asymmetry,
let us see why we fancy and care for a dynamical mechanism accounting
for this asymmetry. We could of course do without one by simply taking
$Y_{B}\approx0.8\times10^{-10}$ as an initial condition, or rather,
a final one. But in this difference lies the whole problem. Indeed,
even if we have no direct cosmological signal dating from before nucleosynthesis,
nothing prevents us from extrapolating the history of the universe
back to temperatures above 200MeV, where a relativistic quark-gluon
plasma is expected. Now the recipe for setting up the required initial
condition in this plasma is the following: count 10 000 000 000 antiquarks;
add to these 10 000 000 0\emph{14} quarks, and start over until the
14 extra quarks pile up to the desired chunk of the universe you want
to build (e.g. $10^{70}$ times to build our single galaxy). Clearly,
even the first step requires a fine tool to achieve the needed $10^{-9}$
relative precision: this is a typical fine tuning problem, equivalent
to the fact that every quark today is the lucky winner of a billion
to one fatal lottery.

Since the far observations mentioned above are in fact only sensitive
to the absolute value $|\eta_{B}|,$ a way to ease this initial fine
tuning problem would be to imagine a baryon separating process that
would leave patches having slightly more baryons, and others having
slightly more anti-baryons, making up a globally symmetric universe
on the whole. However, such a baryon separating process would have
to operate before the epoch where baryons start annihilating, namely
$T_{sep}>20\mathrm{MeV}$, which means a small causal horizon $H^{-1}(T_{sep})<H^{-1}(20\mathrm{MeV})$
and thus a small baryonic number inside it: $B_{causal}<Y_{B}\; s\, H^{-3}|_{20\mathrm{MeV}}\approx10^{-10}(M_{Pl}/20\mathrm{MeV})^{3}\approx10^{52}\approx M_{earth}/m_{\mathrm{p}}$.
This maximal {}``matter island'' size is way too small, given that
NASA missions to other planets survived, or given the arguments about
galactic cosmic rays presented above. Actually, a lower bound on the
size of our matter island in a globally symmetric universe can be
obtained from the $\approx70\mathrm{GeV}$ $\gamma$ rays produced
by annihilations at the island boundaries. This matter island minimal
size turns out to be\cite{Cohen:1998ac} of the order of the present
causal horizon or visible universe.

This explains why we need \emph{baryogenesis}, a global mechanism
operating everywhere before $T=20\mathrm{MeV}$, which can dynamically
transform a symmetric initial condition $Y_{B}=0$ into the observed
$Y_{B}\approx0.8\times10^{-10}\not=0$\textit{,} \textit{\emph{and
thus explain}} \textit{{}``why there is something rather than nothing''}
after $\mathrm{p}-\bar{\mathrm{p}}$ annihilation. The necessary conditions\cite{Sakharov:1967dj}
that any such mechanism has to satisfy, have been spelled out as early
as 1967 by A. Sakharov%
\footnote{It is worth noting that at that time, $B$ conservation had not yet
been ruined by Grand Unification concepts, and that Sakharov immediately
sought connections with the proton lifetime, and with the recently
discovered $CP$ violation in $K_{0}-\bar{K}_{0}$ mixing.%
}.

\begin{description}
\item [S.C.1:~Departure~from~equilibrium]is necessary since at thermal
equilibrium, \[
n_{\mathrm{p}}=\int d^{3}k(e^{-\sqrt{k^{2}+m_{\mathrm{p}}^{2}}/T}+1)^{-1}=n_{\bar{\mathrm{p}}}\]
 because of the $CPT$ relation $m_{\mathrm{p}}=m_{\bar{\mathrm{p}}}$.
Some breakdown of chemical equilibrium is also necessary since otherwise,
microreversibility requires the rates for any process and its inverse
to be equal.
\item [S.C.2:~C~and~CP~violations]are also necessary, as baryon number
is odd under these symmetries. The simplest way to see it is above
$T_{QCD}\approx200$ MeV where baryon number is carried by quarks:
\[
n_{B}=\frac{1}{3}(\underbrace{n_{q_{L}}-n_{\overline{q_{L}}}}_{SU(2)\,\mathrm{doublets}}+\underbrace{n_{q_{R}}-n_{\overline{q_{R}}}}_{SU(2)\,\mathrm{singlets}})\Rightarrow\left\{ \begin{array}{llll}
CP: & q_{L}\leftrightarrow\overline{q_{L}}; & B\leftrightarrow-B\\
C: & q_{L}\leftrightarrow\overline{q_{R}}; & B\leftrightarrow-B\end{array}\right.\]
C violation, like parity, is maximal in the Standard Model and is
thus no problem. CP violation is however also essential, and is responsible
for deciding whether to make baryons or anti-baryons out of the CP
symmetric initial condition $Y_{B}=0$.
\item [S.C.3:~B~violation]is obviously needed to change the baryon number
in a comoving volume.
\end{description}
Reviewing these conditions makes it obvious that baryogenesis \emph{needs}
some particle physics inputs from the micro-world, unlike gamma ray
bursts, supernova explosions, ultra-high energy cosmic rays or other
astroparticle physics puzzles which might ultimately find macroscopic
resolutions. 

In particular, the 3rd baryon violation condition makes a strong appeal
to particle physics beyond the standard model. After the conception
of Grand Unified Theories, baryon number violation had a natural niche
at energies $\approx10^{15}\mathrm{GeV}$, which was then the natural
scale for baryogenesis. However, it was soon realized\cite{'tHooft:1976up}
that, even in the Standard Model, an $SU(2)_{L}$ generalization of
the triangle anomaly responsible for the process $\pi_{0}\to\gamma\gamma$
could violate the conservation of any current of the form: \begin{equation*}
\partial_\mu J^\mu_L\doteq\sum_{i\in  \textrm{doublets}}
\partial_\mu[\bar\psi^i_L e_i\gamma^\mu\psi^i_L]=
(\sum_i e_i)\frac{g_W^2}{16\pi^2}F_{\mu\nu}\tilde F^{\mu\nu}\qquad
\setlength{\unitlength}{1mm} 
\raisebox{-6ex}[1ex]{
\begin{fmfgraph*}(25,20)
\fmfleft{c}
\fmfright{w1,w2}
\fmf{fermion,tension=0.2}{vw2,vc,vw1}
\fmf{fermion,lab.side=left,label=$i$,tension=0.2}{vw1,vw2}
\fmf{dashes,lab.side=left,label=$e_i\gamma^\mu\frac{1-\gamma_5}{2}$}{c,vc}
\fmf{boson,lab.side=right,label=$g_W W_\nu\gamma^\nu\frac{1-\gamma_5}{2}$}{w1,vw1}
\fmf{boson,lab.side=left,label=$g_W W_\rho\gamma^\rho\frac{1-\gamma_5}{2}$}{w2,vw2}
\end{fmfgraph*}}
\qquad
\end{equation*} where $e_{i}$ is any charge assigned to the doublet $i$. In the
presence of \emph{instantons}, \emph{i.e.} non-zero $W$-field solutions
tunneling between different topological vacua and having an integer
instanton number $N=\frac{g_{W}^{2}}{32\pi^{2}}\int d^{4}x\, F\tilde{F}$,
the charge $Q_{L}$ associated with this current changes by $\Delta Q_{L}\doteq\Delta[\int d^{3}xJ_{L}^{0}]=2N\sum_{i}e_{i}$.
So for instance, if we consider the left baryonic charge $Q_{L}=B_{L}$
(which corresponds to the choice $e\equiv0$ except $e_{u_{L}}=e_{d_{L}}=\frac{1}{3}$)
we find a violation proportional to the number of generations $\Delta B_{L}=n_{gen}N$:
\slashchar{B} \emph{does exist} in the SM! Notice that at the same
time, the left lepton number $Q_{L}=L_{L}$ ($e\equiv0$ except $e_{\nu_{L}}=e_{e_{L}}=1$)
changes by the same amount $\Delta L_{L}=n_{gen}N=\Delta B_{L}$,
so that $B-L$ is left intact (as should be the case for a gaugeable
symmetry). The rate of these tunneling anomalous processes $\Gamma_{tunnel}\propto e^{-c{\mygreen N}/g_{W}^{2}}$
is low enough to preserve the proton stability. At high energies or
temperatures\cite{Rubakov:1996vz}, instead of tunneling under the
potential barrier that must be present in the complicated field space,
there exists a possibility to classically roll over the barrier. In
the broken electro-weak phase when $v=\langle H\rangle\not=0$, this
classical rate is governed by a saddle point solution called the \emph{sphaleron},
$\Gamma_{class.over}(T)\propto e^{-E_{sphaleron}/T}\approx e^{-10M_{W}/T}$
which is again very low. In the unbroken phase $v=0$ however, it
was found\cite{Kuzmin:1985mm} to be much faster $\Gamma_{class.over}(T)\propto\alpha_{W}^{5}T^{4}$.
This important result meant that once above the ElectroWeak Phase
Transition $T_{EWPT}\approx100\mathrm{GeV}$, S.C.3's need for baryon
number violation beyond the SM can be evaded. The (nearly) known physics
at the EWPT being the last chance for an operative baryogenesis, this
opened the way to a minimal bottom-up strategy: starting from the
well-established standard model, and adding as little ingredients
as necessary to obtain a successful baryogenesis.

In the SM, S.C.2 is in principle satisfied by the nearly maximal CKM
phase $\delta_{CKM}$, but only manifest itself in delicate quantum
interferences. The required coherence between quarks is likely to
be destroyed in a hot plasma undergoing strong interaction collisions,
which can GIM suppress the final asymmetry by as much as 15 orders
of magnitude\cite{Gavela:1994dt}. In practice, CKM \slashchar{CP}
is insufficient. A second failure of SM baryogenesis concerns S.C.1.
Indeed, the expansion rate of the universe around $T_{EWPT}\approx100\mathrm{GeV}$
is too slow for any relevant SM process to get out of equilibrium.
A first order phase transition, where bubbles of the broken EW symmetry
phase expand in the middle of an unbroken phase plasma, can efficiently
amplify non-equilibrium effects around the critical temperature. However,
in the pure SM, this requires a scalar field mass much lower\cite{Bochkarev:1990fx}
than the present experimental lower bound $m_{h}>114\mathrm{GeV}$.
The baryon asymmetry $Y_{B}\approx10^{-10}$, which seemed too small
from an initial conditions point of view, now seems too large for
SM baryogenesis. 

In the Minimal Supersymmetric extension of the SM, which doesn't need
baryogenesis arguments to be worth considering, both of these problems
can in principle find answers. Indeed, the plethorous scalar fields
that come with Supersymmetry can modify the effective potential of
the theory and reinforce the strength of the EW phase transition (S.C.1),
especially\cite{Carena:1997ki} for light $\tilde{t_{R}}$. Moreover,
the MSSM offers new $CP$ violating phases (S.C.2) less prone to GIM
suppression than the CKM one. Nevertheless, the naturalness of these
ideas has suffered\cite{Cline:2002aa} from the rise of the experimental
lower bound on $m_{h}$, to a point that many consider now extremely
contrived.

\section{Thermal Leptogenesis}

At the same time, the fancy for neutrino oscillations and masses has
solidified into a more and more established fact, and the inclusion
of neutrino masses in the SM changes the shape of our bottom up program.
The reason neutrino masses may have anything to do with baryogenesis
have been recognized by Fukugita and Yanagida\cite{Fukugita:1986hr}
immediately after the discovery\cite{Kuzmin:1985mm} of unsuppressed
anomalous processes in the unbroken EW phase. Indeed, just as elastic
scatterings (which change particle positions but not numbers) will
tend to uniformize the particle density in a box, anomalous processes
(which we saw change $B_{L}$ and $L_{L}$ but not $B_{L}-L_{L}$)
will tend to redistribute an asymmetry carried solely by leptons ($L_{L}=-1,$
$B_{L}=0$) into a more evenly shared asymmetry ($L_{L}=-2/3$, $B_{L}=1/3$).
Since the lepton number is today mostly carried by furtive $T\approx2^{\circ}\mathrm{K}$
neutrinos, the only observable effect of this redistribution is the
generation of a baryon asymmetry. We thus see that the problem of
producing a correct baryon asymmetry is solved if we find a mechanism
to produce a lepton asymmetry $Y_{L_{L}}\approx-3\;10^{-10}$ anytime
before $T_{EWPT}$, \emph{i.e.} if we find a \emph{leptogenesis} mechanism.
Translating Sakharov Conditions to leptogenesis, it is clear that
neutrino masses (and especially Majorana ones%
\footnote{Notice however that without Majorana masses, a clever use of the $L_{L}$
breaking by Dirac masses can also do\cite{Dick:1999je,Murayama:2002je},
even if more contrived.%
}) offer a new way to satisfy S.C.3. Before detailing how the others
conditions can be met, let us explain why we feel leptogenesis can
fit in the minimal bottom up approach we outlined.

As reviewed at length at this conference, neutrino oscillations point
to non vanishing neutrino masses. Barring the introduction of a scalar
$SU(2)$ triplet which alters the $m_{Z}/m_{W}$ ratio, Lorentz invariance
then requires the introduction of $SU(2)$ singlet right-handed neutrino
fields $N$ which carry no gauge charge, and can thus enjoy Majorana
masses a priori disconnected from the EW scale. The most general mass
terms for leptons then read: \begin{equation}
\mathcal{L}_{mass}=\bar{L}H.\frac{1}{v}\diag(m_{e,\mu,\tau}).l_{R}+\frac{1}{2}\bar{N}^{c}.\diag(M_{1,2,3}).N+\bar{L}\tilde{H}.\underbrace{V_{CKM}^{\dagger}.\frac{1}{v}\diag(m_{1,2,3}^{D}).U_{R}}_{{\textstyle \doteq Y_{li}}}.N\label{eq:fullmassL}\end{equation}
 where we have decomposed the 18 parameters of the complex Yukawa
couplings $Y_{li}$ between neutrino flavor $l=e,\mu,\tau$ and Majorana
mass eigenstate $N_{i=1,2,3}$ into: 

\begin{itemize}
\item 3 Dirac mass eigenvalues $m_{1,2,3}^{D}$; 
\item one CKM-like matrix of $SU(3)/U(1)^{4}$, $V_{CKM}(\vec{\theta}_{L},\delta_{L})=V_{23}(\theta_{L}^{23}).V_{13}(\theta_{L}^{13},\delta_{L}).V_{12}(\theta_{L}^{12})$,
containing 3 angles and 1 phase%
\footnote{These parameters in the lepton sector are a priori unrelated to those
in the quark sector, unless imposing a quark-lepton symmetry, natural
for instance in $SO(10)$ GUTs.%
}; 
\item one $SU(3)$ matrix $U_{R}=\diag(e^{i\psi_{R}^{1}},e^{i\psi_{R}^{2}},e^{i\psi_{R}^{3}}).V(\vec{\theta}_{R},\delta_{R}).\diag(e^{i\vec{\phi}_{R}})$
containing 3 angles and 5 phases (both $\psi_{R}^{i}$ and $\phi_{R}^{i}$
separately adding up to 0);
\item 3 phases $\diag(e^{i\vec{\phi}_{CKM}})$ which could multiply $V_{CKM}$
from the right and have been reabsorbed by a common rephasing of $L$
and $l_{R}$.
\end{itemize}
At this stage, the simplest and most natural way of accounting for
the extreme smallness of neutrino masses is to leave Dirac masses
$m^{D}$ around the EW scale where they belong, or at least close
to other Dirac \emph{e.g.} up-quarks masses, and use the fact that
$M_{i}$'s are not a priori related with any scale to raise their
value, which reduces the lightest neutrino masses by the see-saw mechanism\cite{Yanagida:1979xy,Gell-Mann:1980vs,Mohapatra:1980ia}.
Indeed, for energies below $M_{i}$'s, the decoupled $N_{i}$ fields
can be integrated out, leaving an effective mass Lagrangian for the
fields $\bar{L}=(\bar{l}_{L},\bar{\nu})$: \begin{equation}
\mathcal{L}_{mass}\approx\bar{l}_{L}.\diag(m_{e,\mu,\tau}).l_{R}+\frac{1}{2}\bar{\nu}^{c}.\mathcal{M}.\nu\,;\,\mathcal{M}\doteq U_{MNS}.\diag(m_{1,2,3}).U_{MNS}^{T}\approx v^{2}\; Y.M^{-1}.Y^{T}\label{eq:seesaw}\end{equation}
where the light neutrinos mass matrix $\mathcal{M}$ (as promised
inversely proportional to heavy Majorana one $M$) is a symmetric
complex matrix which can again be decomposed into 3 eigenvalues, and
a mixing matrix $U_{MNS}\doteq V(\theta_{atm},\theta_{Chooz},\theta_{sun},\delta).\diag(e^{i\vec{\phi}_{L}})$
containing 3 angles directly measured by oscillations, and only 3
phases (comparing with the decomposition of $U_{R}$, the difference
is that the would-be $\psi_{L}$ phases can be reabsorbed into a common
rephasing of $L$ and $l_{R}$, like $\phi_{CKM}$). It is worth noting
that if we plug $Y$ from equ.~\ref{eq:fullmassL} into the see-saw
formula (\ref{eq:seesaw}), we see that the relation between light
neutrino eigenmasses $m_{1,2,3}$ and heavy ones $M_{1,2,3}$ actually
involves the product $U_{eff}\doteq V_{CKM}.U_{MNS}$ which is closely
related to $U_{R}$ in the sense that if $U_{eff}$ is real or diagonal,
so is $U_{R}$. At this moment, increasing the unforbidden Majorana
mass of the unavoidable right-handed neutrinos thus seems theoretically
the most economical way to account for light neutrino masses. However,
as advocated for a long time by Yanagida, these heavy Majorana neutrinos
may play an interesting role in early cosmology and offer a natural
framework for leptogenesis.

Indeed, S.C.3 is obviously met by the coexistence of decay modes with
opposite lepton numbers, while S.C.2 is satisfied if their rates differ
because of interferences between the tree and one loop amplitudes: 

\begin{center}\begin{tabular}{cccc}
\psset{unit=0.85cm}

\raisebox{1.4cm}{$N_{i}\rightarrow l^{-}H^{+}$:}&
\pspicture(0,0)(3.3,2.6)

\small

\rput[cc]{0}(0.2,1.5){$N_i$}

\psline[linewidth=1pt](0.6,1.5)(2,1.5)

\psline[linewidth=1pt](2,1.5)(3,0.5)

\psline[linewidth=1pt]{<}(2.4,1.1)(2.5,1)

\psline[linewidth=1pt,linestyle=dashed](2,1.5)(3,2.5)

\psline[linewidth=1pt]{>}(2.5,2)(2.6,2.1)

\rput[cc]{0}(3.4,0.5){$l^{-}$}

\rput[cc]{0}(3.4,2.5){$H^{+}$}

\rput[br]{0}(2,1.6){$Y_{li}$}

\endpspicture&
\pspicture(-0.5,0)(4.3,2.6)

\small

\rput[cc]{0}(0.1,1.5){$N_i$}

\rput[br]{0}(1.4,1.6){$Y_{l'i}^*$}

\rput[br]{0}(2.6,2.1){$Y_{l'j}$}

\rput[tr]{0}(2.6,0.9){$Y_{lj}$}

\rput[cl]{0}(2.9,1.5){$N_j$}

\rput[cc]{0}(3.8,0.5){$l^{-}$}

\rput[cc]{0}(3.8,2.5){$H^{+}$}

\psline[linewidth=1pt](0.4,1.5)(1.5,1.5)

\psline[linewidth=1pt,linestyle=dashed](1.5,1.5)(2.5,1)

\psline[linewidth=1pt](2.5,2)(2.5,1)

\psline[linewidth=1pt]{<}(2.8,0.85)(3,0.75)

\psline[linewidth=1pt](2.5,1)(3.5,0.5)

\psline[linewidth=1pt](1.5,1.5)(2.5,2)

\psline[linewidth=1pt,linestyle=dashed](2.5,2)(3.5,2.5)

\psline[linewidth=1pt]{>}(2.8,2.15)(3,2.25)

\endpspicture &
\pspicture(0,0)(4.3,2.6)

\small\rput[cc]{0}(0.1,1.5){$N_i$}

\rput[bc]{0}(1.5,2.2){$l'$}

\rput[br]{0}(1.0,1.6){$Y_{l'i}^*$}

\rput[bl]{0}(2,1.6){$Y_{l'j}$}

\rput[cl]{0}(3,1.4){$Y_{lj}$}

\rput[tc]{0}(2.25,1.2){$N_j$}

\rput[cc]{0}(3.8,0.5){$l^{-}$}

\rput[cc]{0}(3.8,2.5){$H^{+}$}

\psline[linewidth=1pt](0.5,1.5)(1,1.5)

\psarc[linewidth=1pt]{->}(1.5,1.5){0.5}{0}{90}

\psarc[linewidth=1pt](1.5,1.5){0.5}{90}{180}

\psarc[linewidth=1pt,linestyle=dashed]{->}(1.5,1.5){0.5}{180}{270}

\psarc[linewidth=1pt,linestyle=dashed](1.5,1.5){0.5}{270}{360}

\psline[linewidth=1pt](2,1.5)(2.5,1.5)

\psline[linewidth=1pt](2.5,1.5)(3,1)

\psline[linewidth=1pt]{<}(3,1)(3.5,0.5)

\psline[linewidth=1pt,linestyle=dashed]{->}(2.5,1.5)(3,2)

\psline[linewidth=1pt,linestyle=dashed](3,2)(3.5,2.5)

\endpspicture \tabularnewline
 {\small $N_{i}\rightarrow l^{+}H^{-}$: }&
{\small $Y_{li}^{*}$}&
$\sum_{l',j}Y_{l'i}Y_{l'j}^{*}Y_{lj}^{*}$&
$\sum_{l',j}(Y_{l'i}Y_{l'j}^{*}+i\leftrightarrow j)Y_{lj}^{*}$\tabularnewline
\end{tabular}\end{center}

\smallskip{}
\noindent Each decaying $N_{i}$ thus generates a leptonic CP asymmetry
\[
\delta_{i}\doteq\frac{\sum_{l}\Gamma(N_{i}\rightarrow l+H)-\Gamma(N_{i}\rightarrow\overline{l}+H^{\dagger})}{\sum_{l}\Gamma(N_{i}\rightarrow l+H)+\Gamma(N_{i}\rightarrow\overline{l}+H^{\dagger})}\stackrel{M_{i}\ll M_{j}}{\approx}-\frac{3}{16\pi}\frac{\im(A_{ij}^{2})}{A_{ii}}\frac{M_{i}}{M_{j}}\]
 with $A_{ij}=(Y^{\dagger}Y)_{ij}=U_{R}^{\dagger}.\diag(m_{1,2,3}^{D})^{2}.U_{R}$
being the relevant combination of Yukawa couplings, whose diagonal
terms contain the lifetimes $\Gamma_{i}\propto A_{ii}M_{i}$ while
off-diagonal terms carry CP violation if $U_{R}$ (and thus $U_{eff}=V_{CKM}.U_{MNS}$)
is complex. Notice that for $M_{i}\approx M_{j}$, the asymmetry can
be enhanced by the resonant self-energy contribution\cite{Pilaftsis:1998pd,Frere:1999rh}
$\propto1/(M_{j}-M_{i})$ until $\Delta M\approx\Gamma$. 

Turning finally to S.C.1, the decay asymmetry $\delta_{i}$ was here
computed in vacuum, but in a hot plasma, the decay products can recombine
and wash out the asymmetry if the (inverse) decay rate is much larger
than the expansion rate $H$ (S.C.1 fails in local equilibrium) or
equivalently if the dimensionless ratio \[
K_{i}\doteq\Gamma_{N_{i}}(T=M_{i})/H(T=M_{i})\approx1/(1.66\,8\pi\sqrt{g^{*}})\times A_{ii}M_{pl}/M_{i}\]
 is much larger than 1. The lepton asymmetry $Y_{i}$ originating
from species $N_{i}$ in thermal equilibrium is then diluted by a
factor $d\propto1/K$, giving $Y_{i}=\frac{1}{g^{*}}d(K_{i},M_{i})\delta_{i}$.
This lepton asymmetry finally drives anomalous processes to produce
a left baryon asymmetry $Y_{B}\approx-Y_{i}/3$. Assuming as above
that $M_{i}$'s are hierarchical, the contribution from the lightest
$M_{1}$ often dominates in which case the final result to be confronted
with observations takes the simple form:

\[
Y_{B10}\doteq10^{10}\, Y_{B}\approx\frac{10^{10}}{16\pi g^{\star}}d(K_{1},M_{1})\frac{M_{1}}{A_{11}}\sum_{j=2,3}\frac{\im(A_{1j}^{2})}{M_{j}}\approx0.8\]
The dilution factor $d$ is a priori smaller than 1, but a more specific
determination requires the numerical solution of the Boltzmann equations
describing how various particles depart from equilibrium in response
to the universe expansion. The result depends mostly on the ratio
$K_{i}$, which is often rewritten as a dimensionful see-saw like
mass $\tilde{m}_{i}\doteq\frac{v^{2}A_{ii}}{M_{i}}=K_{i}\,\tilde{m}^{*}$,
where the critical value corresponding to $K=1$ is $\tilde{m}^{*}=\sqrt{512g^{*}\pi^{5}/90}\; v^{2}/M_{pl}=1.08\,\,10^{-3}\mathrm{eV}$
in the SM. 

\begin{floatingfigure}[r]{0.50\columnwidth}%
\noindent \includegraphics[%
  bb=135bp 232bp 446bp 456bp,
  clip,
  width=0.50\textwidth]{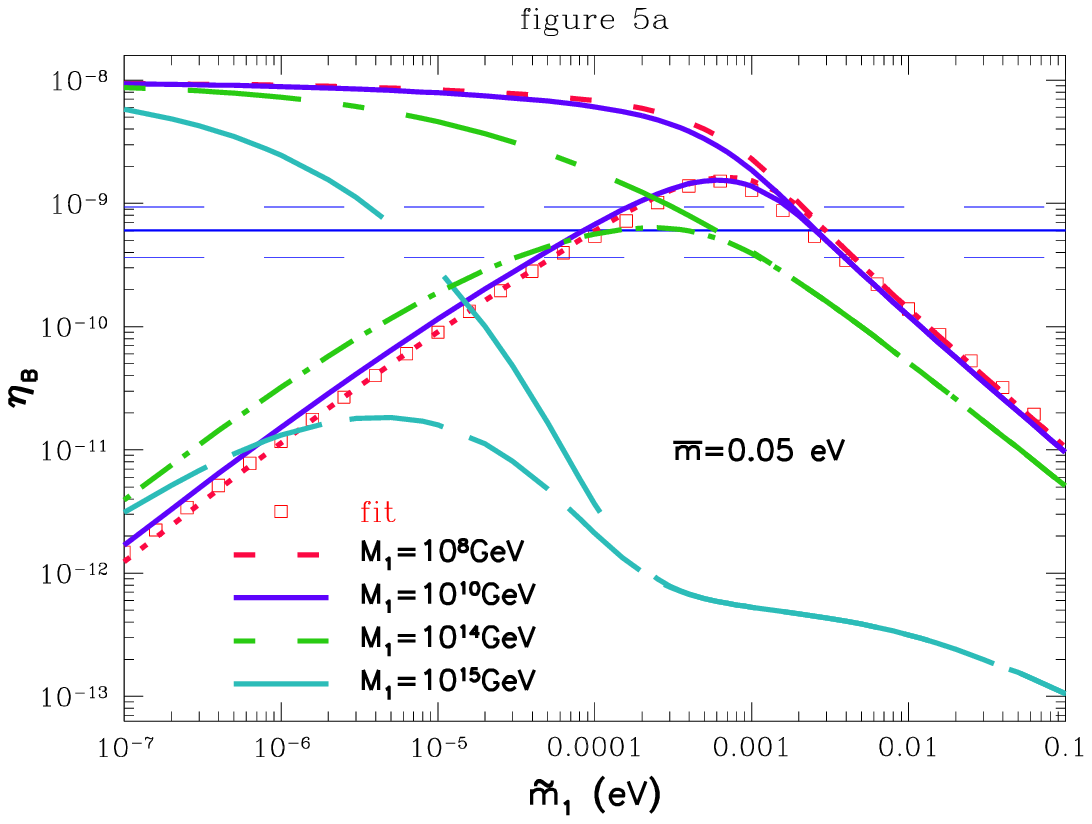}

\caption{\label{cap:dilution}Dependence of $d$ on $\tilde{m}_{1}$ for various
$M_{1}$.}\end{floatingfigure}%
The result is shown\cite{Buchmuller:2002rq} in figure~\ref{cap:dilution}
for a fixed CP asymmetry $\delta_{1}=10^{-6}$, so that $\eta_{B}=10^{-8}d$.
On the right, we see the large $K$ dilution effect from inverse decays
explained above {\small }$d(K>1,M_{1}<10^{15})\propto1/K\propto1/\tilde{m}$.
For small $K$, the curves split in two: the lower one $d(K<1)\propto K$
reflects the difficulty of creating an equilibrium population of right
handed neutrino if their only couplings $Y_{li}$ are too small, and
the upper curve assumes this population is initially present for some
reason. In the first, more natural case, $d$ reaches a maximum of
$d_{max}\approx0.2$ around $K\approx1$. The dependence on $M_{1}$
comes from $2\to2$ scattering effects\cite{Barbieri:1999ma}, whose
relative importance increase at large $M_{1}$ for fixed $m_{\nu},\tilde{m}_{1}$. 

It would be tempting to try and make a direct connection between the
baryon asymmetry $Y_{B10}$ generated by this mechanism and the neutrino
oscillations which make it so natural. It is however impossible if
the Yukawa couplings $Y$'s are totally free. Indeed, for any set
of values $(Y,M)$ transformed by the see-saw into acceptable light
neutrino masses and oscillations, but for which leptogenesis produces
a wrong asymmetry, say $Y_{B10}\ll1$, a simple rescaling increasing
both Yukawas $Y\rightarrow Y^{'}=Y/\sqrt{Y_{B10}}$ and right handed
masses $M\rightarrow M^{'}=M/Y_{B10}$ in such a way as to preserve
the see-saw and thus both light neutrinos and $\tilde{m_{i}}\propto K_{i}$,
would restore%
\footnote{In the approximation where we neglect the $M_{1}$dependence of $d$
above; otherwise, the rescaling factor might take a more complicated
form.%
} a correct asymmetry $Y_{B10}'\approx1$. Similarly, we saw that the
CP phase felt by leptogenesis resides in $U_{R}$ which, through the
see-saw, corresponds to a phase in $U_{eff}=V_{CKM}.U_{MNS}$. Clearly,
the result depends on the assumption for $V_{CKM}$ in the lepton
sector, and short of this assumption, there is no relation possible
with the phases in $U_{MNS}$ which affect oscillations or neutrinoless
double beta decay. 

An interesting upper bound\cite{Hamaguchi:2001gw,Davidson:2002qv}
can however be derived without assuming anything other than hierarchical
right handed masses $M_{i}$. Using a parameterization\cite{Casas:2001sr}
of the Yukawas \[
Y^{\dagger}=v^{-1}\diag(\sqrt{M})R\diag(\sqrt{m^{D}})U_{MNS}^{\dagger}\]
 with a complex orthogonal matrix $R$, one can write \[
\delta_{1}=\frac{-3}{8\pi}\frac{M_{1}}{v^{2}}\frac{1}{A_{11}}\im(Y^{\dagger}\mathcal{M}Y^{*})_{11}=\frac{-3}{8\pi}\frac{M_{1}}{v^{2}}\frac{\sum_{j}m_{j}^{2}\im(R_{1j}^{2})}{\sum_{j}m_{j}|R_{1j}|^{2}}\]
whose maximization over $R$ gives \begin{equation}
|\delta_{1}|\leq\frac{3}{8\pi}\frac{M_{1}}{v^{2}}(m_{3}-m_{1})\label{eq:deltabound}\end{equation}
 Since the dilution factor $d$ cannot exceed $0.2$, the requested
$Y_{B}$ translates into a lower bound on $M_{1}>\frac{0.06\mathrm{eV}}{m_{3}-m_{1}}10^{9}\mathrm{GeV}$.
This is turn puts a lower bound on the reheating temperature after
inflation $T_{reh}>10^{8\rightarrow10}\mathrm{GeV}$. In SUSY models,
the overproduction of unstable gravitinos potentially destroying nucleosynthesis
require $T_{reh,SUSY}<10^{9\rightarrow12}\mathrm{GeV}$, which seems
uncomfortably close to the previous number. A possible loophole is
to relax the hierarchy requirement under which the bound (\ref{eq:deltabound})
was obtained, and let $M_{2}$ get close enough to $M_{1}$ to benefit
from the self-energy diagram enhancement\cite{Ellis:2002eh}.

\begin{figure}
 \psfrag{-7}[c][c]{-7} \psfrag{-6}[c][c]{-6} \psfrag{-5}[c][c]{-5} \psfrag{-4}[c][c]{-4} \psfrag{-3}[c][c]{-3} \psfrag{-2}[c][c]{-2} \psfrag{-1}[c][c]{-1} \psfrag{0}[c][c]{0} \psfrag{m1}[t][b]{\(Log_{10} [m_1/{\textrm{eV}}]\)} \psfrag{4}[r][r]{4} \psfrag{6}[r][r]{6} \psfrag{8}[r][r]{8} \psfrag{10}[r][r]{10} \psfrag{12}[r][r]{12} \psfrag{14}[r][r]{14} \psfrag{16}[r][r]{16} \psfrag{MR}[c][l][1][90]{\(Log_{10} [M_R]\)} \psfrag{VAC: Right handed masses}[c][c]{Vacuum: Right handed masses} \psfrag{se3=0}[tl][l]{{\myblue \(s_{e3}=0\)}} \psfrag{se3=-0.16}[tl][l]{{\myred \(s_{e3}=-0.16\)}} \psfrag{Ueff=1}[l][tl]{{\mygreen \(U_{eff}=1\)}} \psfrag{M1}[r][r]{\(M_1\)} \psfrag{M2}[r][r]{\(M_2\)} \psfrag{M3}[r][r]{\(M_3\)}\includegraphics[%
  width=0.45\textwidth]{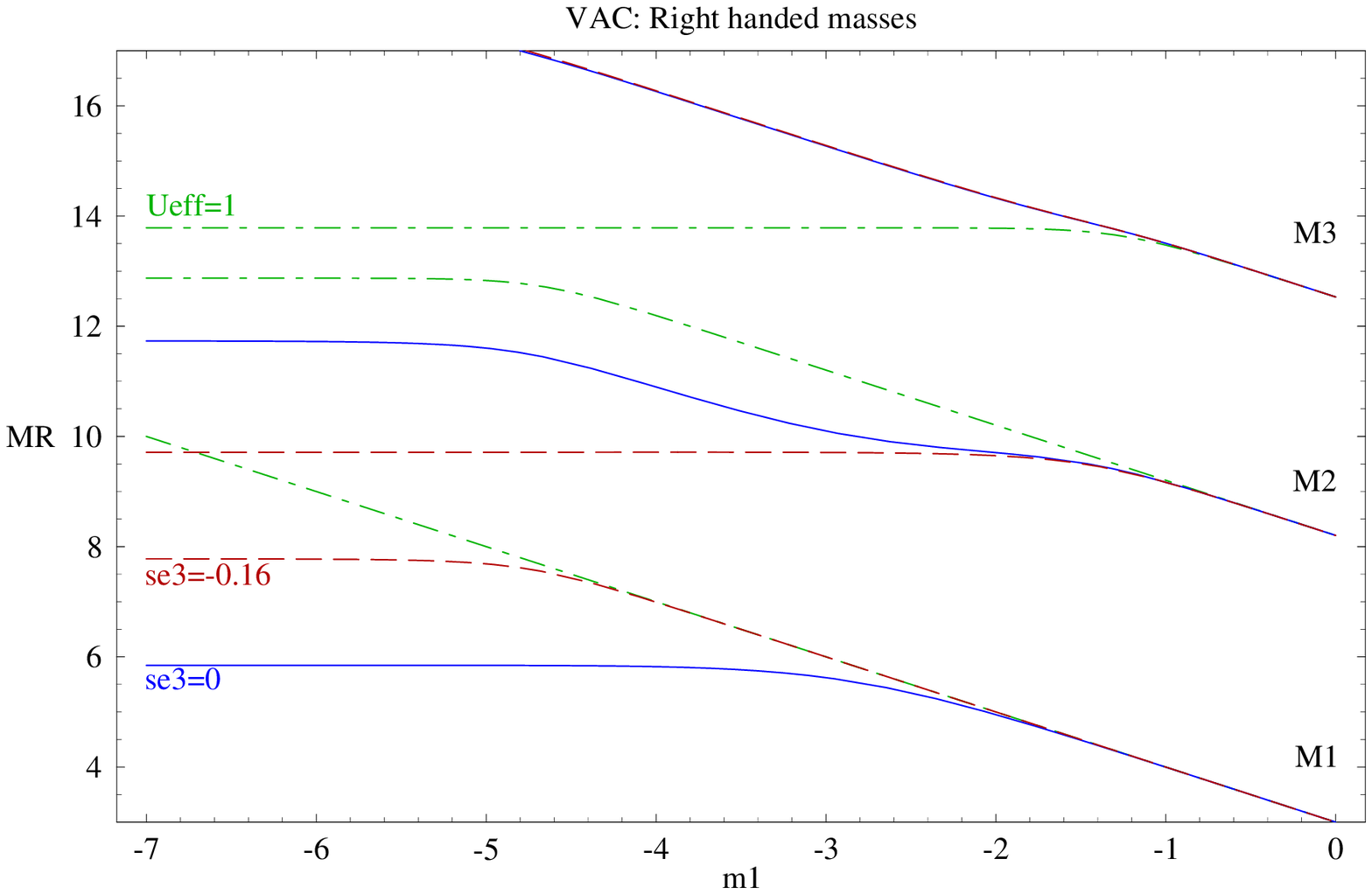} \psfrag{-6}[c][c]{-6} \psfrag{-5}[c][c]{-5} \psfrag{-4}[c][c]{-4} \psfrag{-3}[c][c]{-3} \psfrag{-2}[c][c]{-2} \psfrag{-1}[c][c]{-1} \psfrag{0}[c][c]{0} \psfrag{Log[m1/eV]}[tc][b]{\(Log_{10}[m_1/{\textrm{eV}}]\)} \psfrag{-8}[r][r]{-8} \psfrag{-6}[r][r]{-6} \psfrag{-4}[r][r]{-4} \psfrag{-2}[r][r]{-2} \psfrag{0}[r][r]{0} \psfrag{Log[YB10]}[c][c][1][90]{\(Log_{10}[Y_{B10}]\)} \psfrag{VAC: se3=0}[l][l]{VAC: \(s_{e3}=0\)} \psfrag{VAC: se3=-0.16}[l][l]{VAC: \(s_{e3}=-0.16\)} \psfrag{LOW: se3=-0.16}[l][lb]{LOW: \(s_{e3}=-0.16\)}\includegraphics[%
  width=0.50\textwidth]{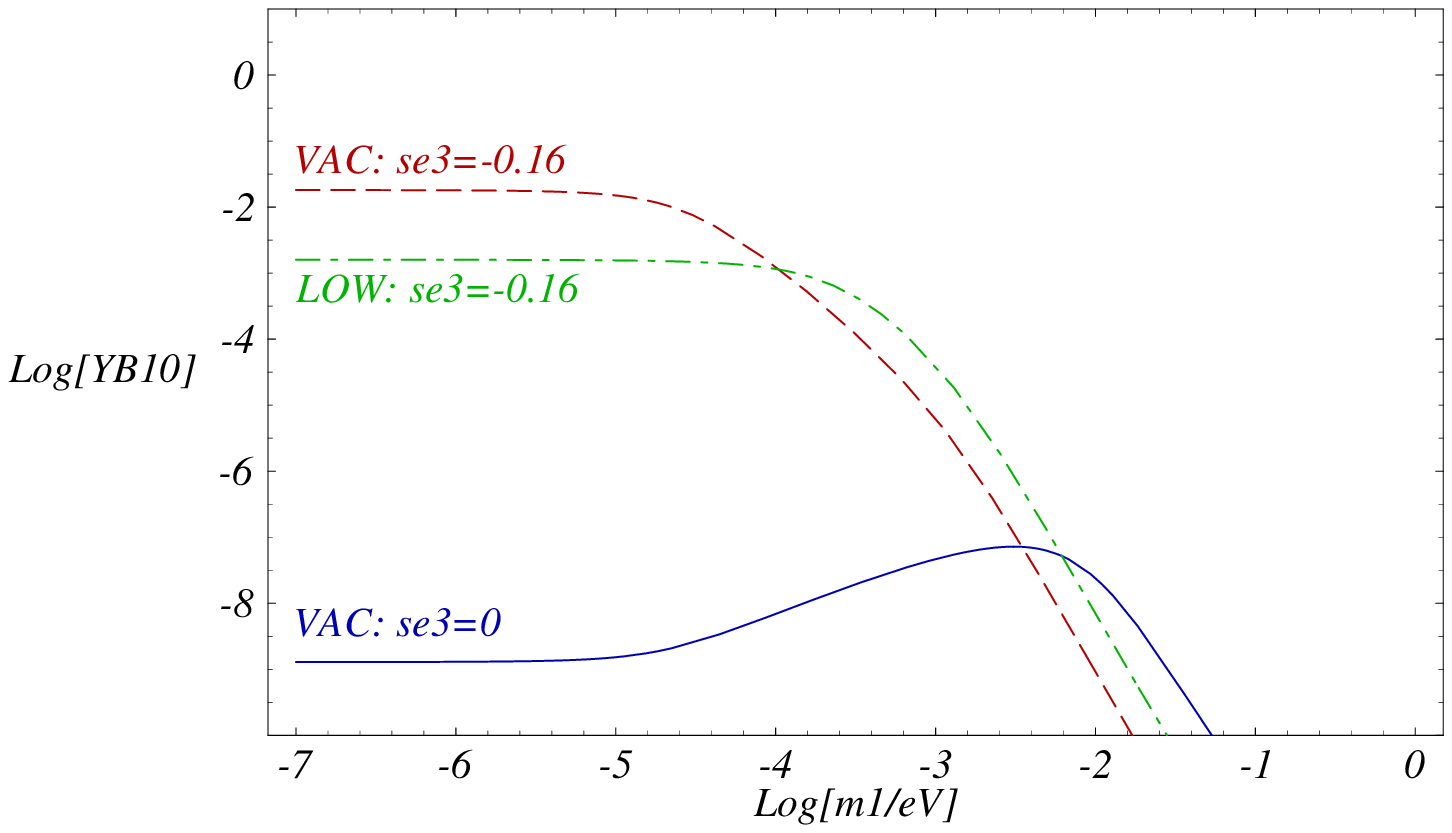}

\caption{\label{cap:SO10}Right handed neutrino masses $M_{i}$ (left) and
insufficient maximal asymmetry (right) in an $SO(10)$-inspired model.}
\end{figure}
To illustrate the power of this bound~(\ref{eq:deltabound}), let
us consider a simple $SO(10)$-inspired example\cite{Nezri:2000pb}
where we extend $b-\tau$ unification and fix Yukawas $Y$ by $m_{\nu}^{D}=m_{u}/3$.
By the see-saw, $M_{i}$ and $U_{R}$ are then determined from the
light neutrino mass matrix $\mathcal{M}$, up to the lightest neutrino
mass $m_{1}$, the MNS element $|U_{e3}|$ and 5 phases which are
taken as free parameters. Right-handed masses turn out hierarchical
(like the assumed Dirac masses), and $M_{1}$ can hardly exceed $10^{8}\mathrm{GeV}$.
The maximal asymmetry is nearly 2 orders of magnitude too small, as
should follow from (\ref{eq:deltabound}). Yet this maximal result
is obtained for a special value of $|U_{e3}|\approx0.16$ which approximately
cancels CKM Cabbibo mixing and allows for the largest $M_{1}$(as
seen from the extreme case $U_{eff}=1$ on the left figure~\ref{cap:SO10}).
All results on figure~\ref{cap:SO10} are for the now disfavored
{}``vacuum'' solar neutrino oscillations ($\Delta m_{sol}^{2}=4.6\,10^{-10}\mathrm{eV^{2}}$),
the LMA solution giving similar but yet smaller $M_{1}$ and asymmetries\cite{Nezri:2000pb}.

\section{Conclusion}

In this short review, we hope to have convinced that the baryon asymmetry
of the universe is rather well established and constitutes an important
particle physics question worth the considerable amount of work it
has attracted. Despite being a single number out of a non repeatable
experiment, it can be seen as one of the only evidences for an incompleteness
of the Standard Model. As to the direction this incompleteness points
at, thermal leptogenesis relying on the heavy right-handed neutrinos
required by the see-saw mechanism, constitutes in our mind a very
suggestive minimal, predictive and not excluded solution. Obviously,
this opinion is shared by the large number of authors who recently
contributed to this field, many of whom may justly feel misrepresented
in this review by lack of space. It would be extremely nice to cross-check
the existence of heavy right-handed neutrinos by some independent
measurement. Unfortunately, apart from lepton flavor violating effects
which might show up in certain cases discussed at this conference,
this will not be easy. Meanwhile, thanks to the upper bound on CP
violation in right-handed neutrino decays, imposing successful leptogenesis
takes an non-trivial slice in the see-saw parameter space which usually
cuts more than simply one dimension. This might shed useful light
in the quest for some order in the present mass anarchy. Finally,
the CP violation needed for leptogenesis cannot simply at present
be related to the one measurable in future long baselines neutrino
experiments, and even less to the one already measured in the quark
sector, unless some relations (like the $SO(10)$-inspired ones discussed
above) are imposed: if we have all the phenomenology needed to account
for CP violation, a real theory is still badly lacking.

\end{fmffile}

\bibliographystyle{/Users/orloff/Documents/paps/lepto/talks/moriond/joproc-hepph}
\bibliography{/Users/orloff/Documents/paps/lepto/talks/moriond/joproc,/Users/orloff/Documents/paps/lepto/leptosc-6}

\end{document}